\documentclass{PoS}
\usepackage{amsmath}

\title{Update on the 2+1+1 flavor QCD equation of state with HISQ}

\ShortTitle{Update on the 2+1+1 flavor QCD equation of state with HISQ}

\author{%
        \speaker{A.~Bazavov}$^{,a,\S}$\thanks{Present address: 
        Department of Physics and Astronomy, University of Iowa, Iowa City, IA 52245, USA},
        C.~Bernard$^b$,
        C.~DeTar$^c$,
        J.~Foley$^c$,
        Steven Gottlieb$^d$,
        Urs M.~Heller$^e$,
        J.E.~Hetrick$^f$,
        J.~Laiho$^g$,
        L.~Levkova$^c$,
        J.~Osborn$^h$,
        R.~Sugar$^i$,
        D.~Toussaint$^j$, ~~~~~
        R.S.~Van de Water$^k$,
        and R.~Zhou$^k$ \\ \\
        \llap{$^a$}Department of Physics, Brookhaven National Laboratory, Upton, NY 11973, USA\\
        \llap{$^b$}Department of Physics, Washington University, St. Louis, MO 63130, USA\\
        \llap{$^c$}Department of Physics and Astronomy, University of Utah, Salt Lake City, UT 84112, USA\\
        \llap{$^d$}Department of Physics, Indiana University, Bloomington, IN 47405, USA\\
        \llap{$^e$}American Physical Society, One Research Road, Ridge, NY 11961, USA\\
        \llap{$^f$}Physics Department, University of the Pacific, Stockton, CA 95211, USA\\
        \llap{$^g$}Department of Physics, Syracuse University, Syracuse, NY 13244, USA\\
        \llap{$^h$}Argonne Leadership Computing Facility, Argonne National Laboratory, Argonne, IL 60439, USA\\
        \llap{$^i$}Department of Physics, University of California, Santa Barbara, CA 93106, USA\\
        \llap{$^j$}Department of Physics, University of Arizona, Tucson, AZ 85721, USA\\
        \llap{$^k$}Theoretical Physics Department, Fermi National Accelerator Laboratory, Batavia, IL 60510, USA\\
        \\
        \llap{$^\S$}E-mail: \email{obazavov@quark.phy.bnl.gov}
}

\abstract{We present recent results on the QCD equation of state 
with 2+1+1 flavors of highly improved staggered quarks (HISQ). We focus 
on three sets of ensembles with temporal extent $N_\tau=6$, $8$ and $10$,
that reach up to temperatures of $967$, $725$ and $580$~MeV, respectively.
The strange and charm quark masses are tuned to the physical values and
the light quarks mass is set to one fifth of the strange.
This corresponds to a Goldstone pion of about $300$~MeV.}

\FullConference{31st International Symposium on Lattice Field Theory LATTICE 2013\\
                 July 29 -- August 3, 2013\\
                 Mainz, Germany}

\begin{document}

\section{Introduction}

Properties of the high-temperature phase of QCD, the quark-gluon plasma, are currently
a subject of investigation in ultra-relativistic heavy-ion collision experiments at RHIC (BNL),
LHC (CERN) and planned future experiments FAIR (GSI) and NICA (JINR). The QCD equation
of state with 2+1 flavors of quarks has been and is being extensively studied on the lattice
\cite{Bazavov:2009zn,Borsanyi:2013bia}
and some preliminary results for the 2+1+1 flavor equation of state (\textit{i.e.}
with a dynamical charm quark) are also available \cite{Borsanyi:2012vn,Bazavov:2012kf}.
Here we report on the continuation of the study initiated in Ref.~\cite{Bazavov:2012kf}.

In heavy-ion experiments heavy quarks are absent in the
colliding nuclei and are created at early stages of the collision. Therefore they play an
important role, both theoretically and experimentally, as probes of the deconfined medium.
Although the charm quark mass is on the order of 10$T_c$ ($T_c=154(9)$~MeV being the chiral
crossover temperature \cite{Bazavov:2011nk}), perturbative \cite{Laine:2006cp} 
and quenched charm lattice \cite{Cheng:2007wu}
calculations indicate that the charm contribution to the equation of state becomes non-negligible
at temperatures as low as 2-3$T_c$. These temperatures are within reach of the heavy-ion
program at LHC. Therefore, it seems timely to include the dynamical charm quark in \textit{ab initio}
QCD calculations.

\section{Lattice setup}

Calculation of the equation of state on the lattice is computationally expensive
because it requires subtraction of ultra-violet divergences, and both finite- and
zero-temperature ensembles with large statistics are needed at every value of the
gauge coupling. Therefore this study has been done along the line of constant physics (LCP)
with the light quark mass set to $m_l=m_s/5$ and makes use of a set of the existing MILC
zero-temperature ensembles \cite{Bazavov:2010ru} on this LCP.

We use the tadpole one-loop improved gauge action and the highly improved staggered
quark (HISQ) action \cite{Follana:2006rc}.
The HISQ action suppresses the taste-exchange interactions
present in the staggered formalism and significantly reduces the mass splittings
between various pion tastes. This feature improves the approach to the continuum
limit at low temperatures. It is also $O(a^2)$-improved, with the Naik (three-link) term,
which controls scaling at high temperatures. In lattice units the charm quark
mass $am_c \sim O(1)$, therefore the Naik term includes a mass-dependent correction, 
$\epsilon_N$~\cite{Follana:2006rc}.
It is derived perturbatively to reproduce the correct charm quark dispersion relation
up to $O((am_c)^4)$.
To set the lattice spacing $a$ we use the scale
$r_1\simeq 0.31$~fm~\cite{Sommer:1993ce}. The strange
and charm quark masses are tuned to the physical values by using the $\pi$, $K$,
$\eta_c$ and $J/\psi$ masses. The tadpole factor defined from the trace of the
plaquette $u_0=\left\langle{\rm Tr} U_p/3\right\rangle^{1/4}$ is determined during
the equilibration of the zero-temperature ensembles. We used finite-temperature
lattices with aspect ratio of four and temporal extent $N_\tau=6$, $8$ and $10$.
(In this round we did not pursue the exploratory $N_\tau=12$ 
ensembles reported in Ref.~\cite{Bazavov:2012kf}, since
to get a reliable signal would require significant computational resources.)
The temperature is set as $T=1/(aN_\tau)$.

We determined the $\beta$-functions by fitting the data to the following Ans\"{a}tze.
For the lattice spacing:
\begin{equation}
\frac{r_1}{a}(\beta) = \frac{c_r^{(0)} f(\beta) + c_r^{(2)} (10/\beta) f^3(\beta)}
  {1 + d_r^{(2)} (10/\beta) f^2(\beta)} \, ,
\label{eq:ar1_para}
\end{equation}
and for the strange and charm quark masses:
\begin{equation}
am_q(\beta) = \frac{c_q^{(0)} f(\beta) + c_q^{(2)} (10/\beta) f^3(\beta)}
  {1 + d_q^{(2)} (10/\beta) f^2(\beta)} \,
\left(\frac{20b_0}{\beta}\right)^{4/9}
\label{eq:am_para}
\end{equation}
where $q=s,c$ and
\begin{equation}
f(\beta) = \left(\frac{10b_0}{\beta}\right)^{-b_1/(2b^2_0)}
 \exp{(-\beta/20b_0)}
\label{eq:beta_fn}
\end{equation}
is the perturbative two-loop $\beta$-function for three flavors.
We checked that using the four-flavor $\beta$-function renormalizes
the coefficients but produces the same (within the numerical
accuracy) results for the scale and quark masses.

The parameters and accumulated statistics for the zero-temperature
ensembles are shown in Table~\ref{tab:LCP_latts}.
The entries marked with asterisk are used to set the LCP and lattice scale.
Corresponding temperatures and the statistics for the finite-temperature
ensembles are shown in Table~\ref{tab_T}.
\begin{table}
\begin{center}
\begin{tabular}{|l|l|l|l|l|l||l|r|}
\hline
$\beta$ & $am_s$ & $am_c$ & $u_0$ & $\epsilon_N$ & $a$, fm & size & TU/$10^3$ \\
\hline
5.400* & 0.091  & 1.339  & 0.83496  & -0.7995 & 0.220 & $16^3 \times 40$  &  5.0 \\
5.469  & 0.0928 & 1.263  & 0.838768 & -0.6905 & 0.206 & $24^3 \times 32$  &  9.7 \\
5.541  & 0.0859 & 1.157  & 0.842646 & -0.5797 & 0.192 & $24^3 \times 32$  &  9.7 \\
5.600* & 0.0785 & 1.080  & 0.845768 & -0.5168 & 0.181 & $16^3 \times 48$  & 12.5 \\
5.663  & 0.0753 & 0.996  & 0.848919 & -0.4571 & 0.170 & $24^3 \times 32$  &  9.7 \\
5.732  & 0.0697 & 0.913  & 0.852242 & -0.4024 & 0.159 & $32^4$            &  3.8 \\
5.800* & 0.065  & 0.838  & 0.85535  & -0.3582 & 0.151 & $16^3 \times 48$  & 98.8 \\
5.855  & 0.0608 & 0.782  & 0.857786 & -0.3195 & 0.140 & $32^4$            & 12.1 \\
5.925  & 0.0561 & 0.716  & 0.860718 & -0.2784 & 0.130 & $32^4$            & 14.5 \\
6.000* & 0.0509 & 0.635  & 0.86372  & -0.2308 & 0.121 & $24^3 \times 64$  & 11.4 \\
6.060  & 0.0481 & 0.603  & 0.865978 & -0.2010 & 0.113 & $32^4$            &  9.7 \\
6.122  & 0.0448 & 0.558  & 0.86824  & -0.1838 & 0.106 & $32^4$            &  9.7 \\
6.180  & 0.042  & 0.518  & 0.870236 & -0.1613 & 0.100 & $32^4$            &  7.8 \\
6.238  & 0.0392 & 0.482  & 0.872177 & -0.1418 & 0.094 & $32^4$            &  9.7 \\
6.300* & 0.037  & 0.440  & 0.874164 & -0.1204 & 0.089 & $32^3 \times 96$  &  6.0 \\
6.530  & 0.028  & 0.338  & 0.880888 & -0.0734 & 0.070 & $36^3 \times 48$  &  2.8 \\
6.720* & 0.024  & 0.286  & 0.885773 & -0.0533 & 0.058 & $48^3 \times 144$ &  5.9 \\
7.000* & 0.0158 & 0.188  & 0.892186 & -0.0235 & 0.045 & $64^3 \times 192$ &  0.7 \\
7.140  & 0.0145 & 0.172  & 0.895074 & -0.0197 & 0.039 & $64^3 \times 72$  &  1.3 \\
7.285  & 0.0124 & 0.148  & 0.89789  & -0.0146 & 0.034 & $64^3 \times 96$  &  0.9 \\
\hline
\end{tabular}
\end{center}
\caption{The parameters of the HISQ ensembles
along the $m_l=m_s/5$ LCP: inverse gauge coupling $\beta=10/g^2$,
input strange and charm quark masses, tadpole factor $u_0$, the Naik term correction
$\epsilon_N$, approximate lattice spacing $a$, lattice volume and the statistics
in thousands of molecular dynamics time units.
}
\label{tab:LCP_latts}
\end{table}

\begin{table}
\centering
\begin{tabular}{|l|l|r|l|r|l|r|l|r|}
\hline
 & \multicolumn{2}{|c|}{$N_\tau=6$} & \multicolumn{2}{|c|}{$N_\tau=8$} &
 \multicolumn{2}{|c|}{$N_\tau=10$}  \\
\hline
$\beta$ & $T$ & TU/$10^3$ & $T$ & TU/$10^3$ & $T$ & TU/$10^3$  \\
\hline
5.400 & 149 & 10  &     &     &     &      \\
5.469 & 160 & 34  &     &     &     &      \\
5.541 & 171 & 34  &     &     &     &      \\
5.600 & 182 & 10  & 136 & 20  &     &      \\
5.663 & 193 & 34  & 145 & 30  &     &      \\
5.732 & 207 & 30  & 155 & 42  &     &      \\
5.800 & 218 & 10  & 163 & 10  & 131 & 40  \\
5.855 & 235 & 30  & 176 & 45  & 140 & 42  \\
5.925 & 253 & 30  & 190 & 45  & 152 & 42  \\
6.000 & 272 & 10  & 204 & 10  & 163 & 40  \\
6.060 & 291 & 30  & 218 & 39  & 175 & 42  \\
6.122 & 310 & 30  & 233 & 39  & 186 & 42  \\
6.180 &     &     & 247 & 11  & 197 & 40  \\
6.238 &     &     &     &     & 210 & 14  \\
6.300 & 369 & 10  & 277 & 10  & 222 & 10  \\
6.530 &     &     & 352 & 11  & 282 &  9  \\
6.720 & 567 & 10  & 425 & 10  & 340 & 10  \\
7.000 &     &     & 548 &  8  & 438 & 20  \\
7.140 & 843 &  7  & 632 & 11  & 506 & 19  \\
7.285 & 967 &  3  & 725 & 11  & 580 & 17  \\
\hline
\end{tabular}
\caption{Inverse gauge coupling, temperatures and the statistics
in thousands of molecular dynamics time units for the
HISQ finite-temperature ensembles along the $m_l=m_s/5$ LCP.}
\label{tab_T}
\end{table}

\section{Trace anomaly}

At lattice spacing $a$ the trace of the 
energy-momentum tensor, or interaction measure, can be related
to the partition function as
\begin{equation}
\varepsilon-3p = -\frac{T}{V}\frac{d\ln Z}{d\ln a},\,\,\,\,\,\,\,\,\,\,\,\,\,
Z=\int DU D\bar\psi D\psi \exp\left(-S_{g}-S_{f}\right),
\end{equation}
where $\varepsilon$ is the energy density and $p$ is the pressure.

To normalize the trace anomaly to zero at zero temperature one can take a
difference between the finite- and zero-temperature observables at the
same values of the gauge coupling and quark masses, \textit{i.e.} for
an observable $X$:
\begin{equation}
  \Delta(X)=\langle X \rangle_\tau - \langle X \rangle_0\,.
\end{equation}
The trace anomaly is then given in terms of the basic observables that enter
into the action:
\begin{eqnarray}
  \frac{\varepsilon-3p}{T^4}&=&-R_\beta(\beta)\left[\Delta(S_{g})+
  R_u(\beta)\Delta\left(\frac{dS_g}{du_0}\right)\right]
  + R_\beta(\beta) R_{m_s}(\beta)
  \left[2m_l\Delta(\bar\psi_l\psi_l)+m_s\Delta(\bar\psi_s\psi_s)\right]\nonumber\\
  &+&R_\beta(\beta) R_{m_c}(\beta)
  \left[m_c\Delta(\bar\psi_c\psi_c)+
  R_{\varepsilon_N}(\beta)\Delta\left(\bar\psi_c
  \left[\frac{dM_c}{d\varepsilon_N}\right]\psi_c\right)\right]\,.\label{e3p_all}
\end{eqnarray}
The change of the lattice spacing and the parameters of the action along the LCP
are controlled by the $\beta$-functions:
\begin{eqnarray}
R_\beta(\beta) &=& T \frac{{\rm d} \beta}{{\rm d}T} =
 - a \frac{{\rm d} \beta}{{\rm d}a} = (r_1/a)(\beta) \,
 \left( \frac{{\rm d} (r_1/a)(\beta)}{{\rm d}\beta} \right)^{-1} \, ,
\label{eq:Rb}\\
R_{m_q}(\beta) &=& \frac{1}{am_q(\beta)} \frac{{\rm d} am_q(\beta)}{{\rm d}\beta}
\qquad \text{for} ~ q=s, c \, ,
\label{eq:Rm}\\
R_u(\beta) &=& \beta \frac{{\rm d} u_0(\beta)}{{\rm d}\beta}
\, , \\
R_{\epsilon}(\beta) &=& \frac{{\rm d} \epsilon_N(\beta)}{{\rm d}\beta} \, .
\end{eqnarray}
As function of $\beta$ the tadpole factor $u_0$ is fit to $u_0(\beta)=c_1+c_2e^{-d_1\beta}$, and
the Naik term correction $\epsilon_N$ to a polynomial in $\beta$.

\begin{figure}
\begin{center}
\includegraphics[width=0.495\textwidth]{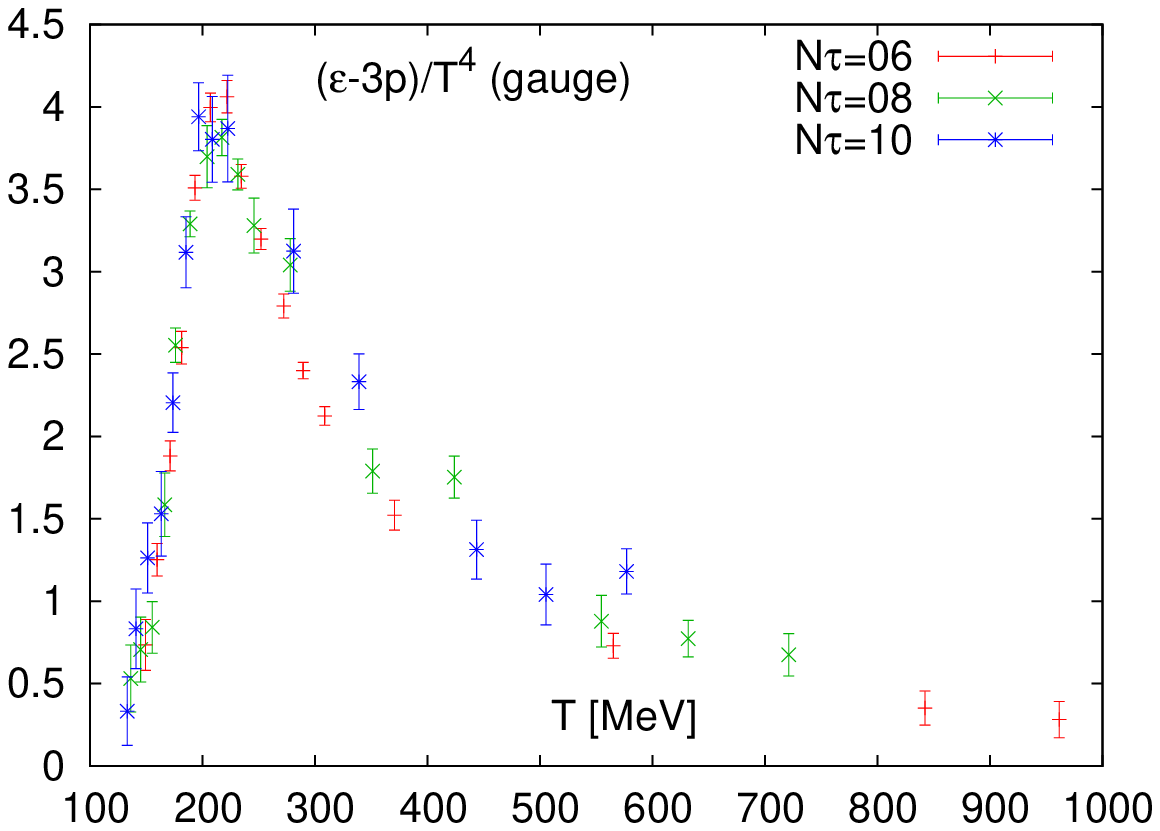}\hfill
\includegraphics[width=0.495\textwidth]{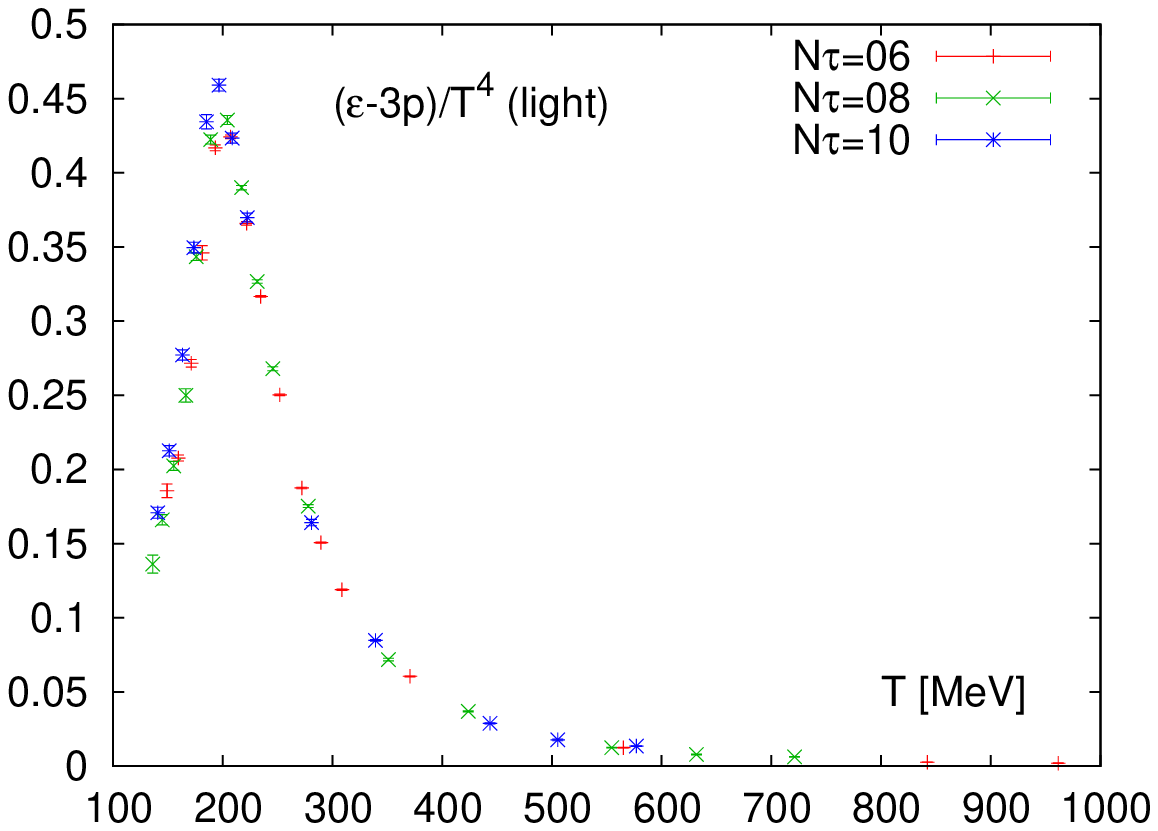}
\includegraphics[width=0.495\textwidth]{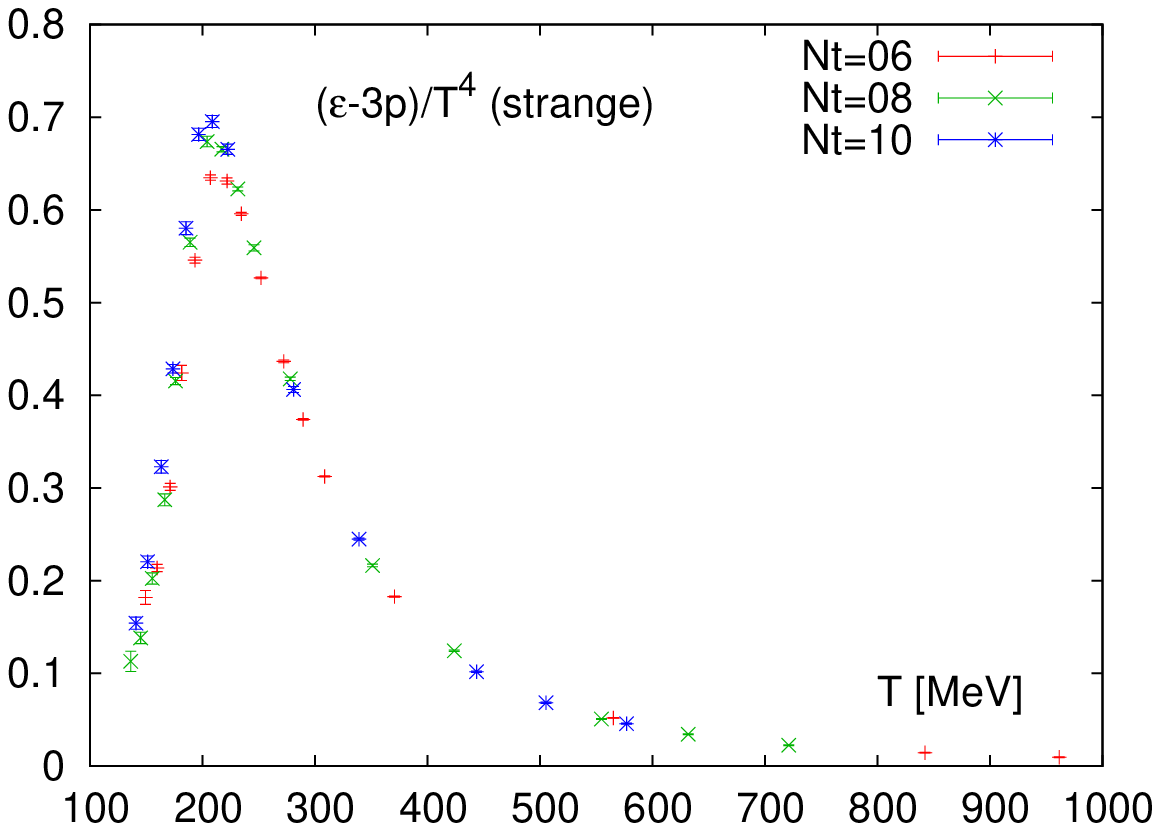}\hfill
\includegraphics[width=0.495\textwidth]{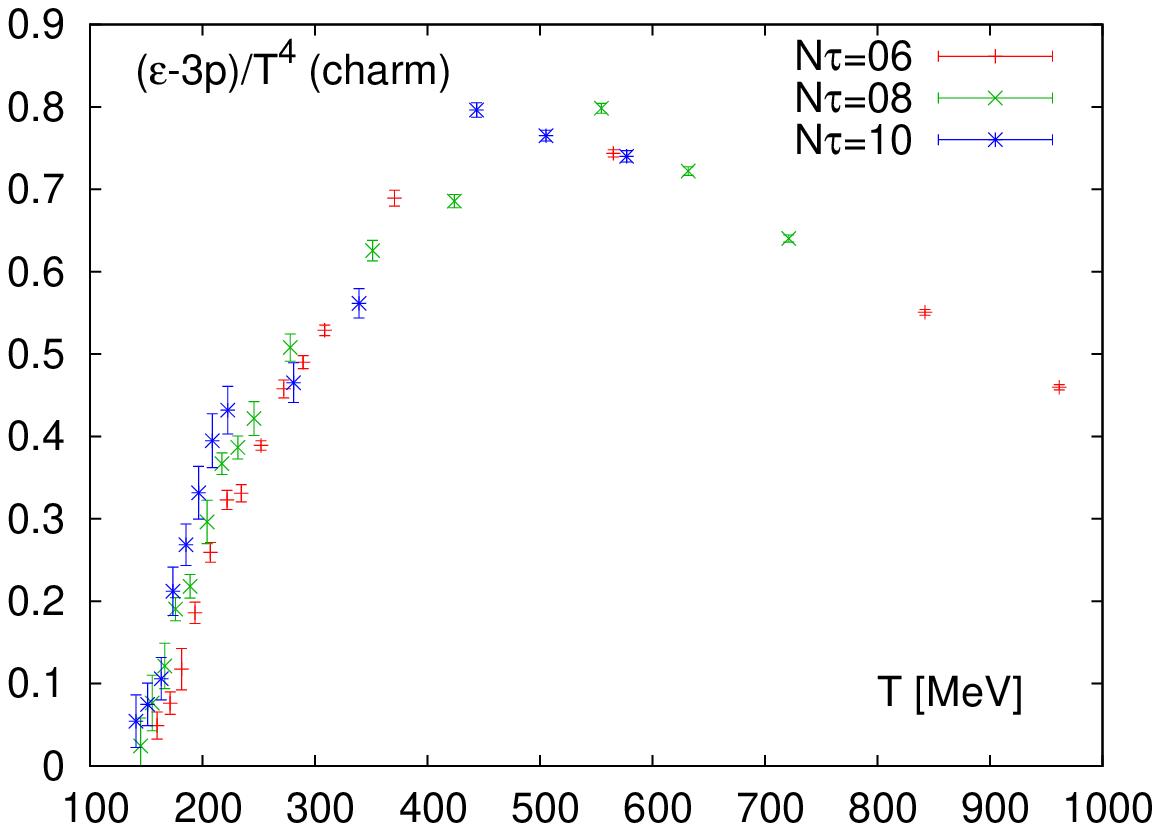}
\end{center}
\vspace{-7mm}
\caption{Contributions to the trace anomaly from the gauge field (top left),
valence light (top right), strange (bottom left) and charm (bottom right)
quarks.}
\vspace{-3mm}
\label{fig_Sgf}
\end{figure}

The first term in Eq.~(\ref{e3p_all}) describes the contribution to the trace
anomaly from the gauge field, the second from the (valence)
light and strange quarks and the third from the (valence) charm quark. These
four quantities are shown in Fig.~\ref{fig_Sgf}. In the peak region around 200~MeV
the trace anomaly is dominated by the gauge part. Light and strange quark 
contributions also reach their maxima around that temperature, while the charm part
at 200~MeV contributes less than 10\% to the total result. The charm contribution
reaches its maximum around 500~MeV and accounts for 40\% of the value of the
trace anomaly at that temperature. (Note, that this is the dominant, valence,
charm quark contribution, while the effect of the sea charm can only be quantified by
comparing to the 2+1 flavor equation of state. Such an analysis is left for the future.)

\begin{figure}
\begin{center}
\includegraphics[width=0.495\textwidth]{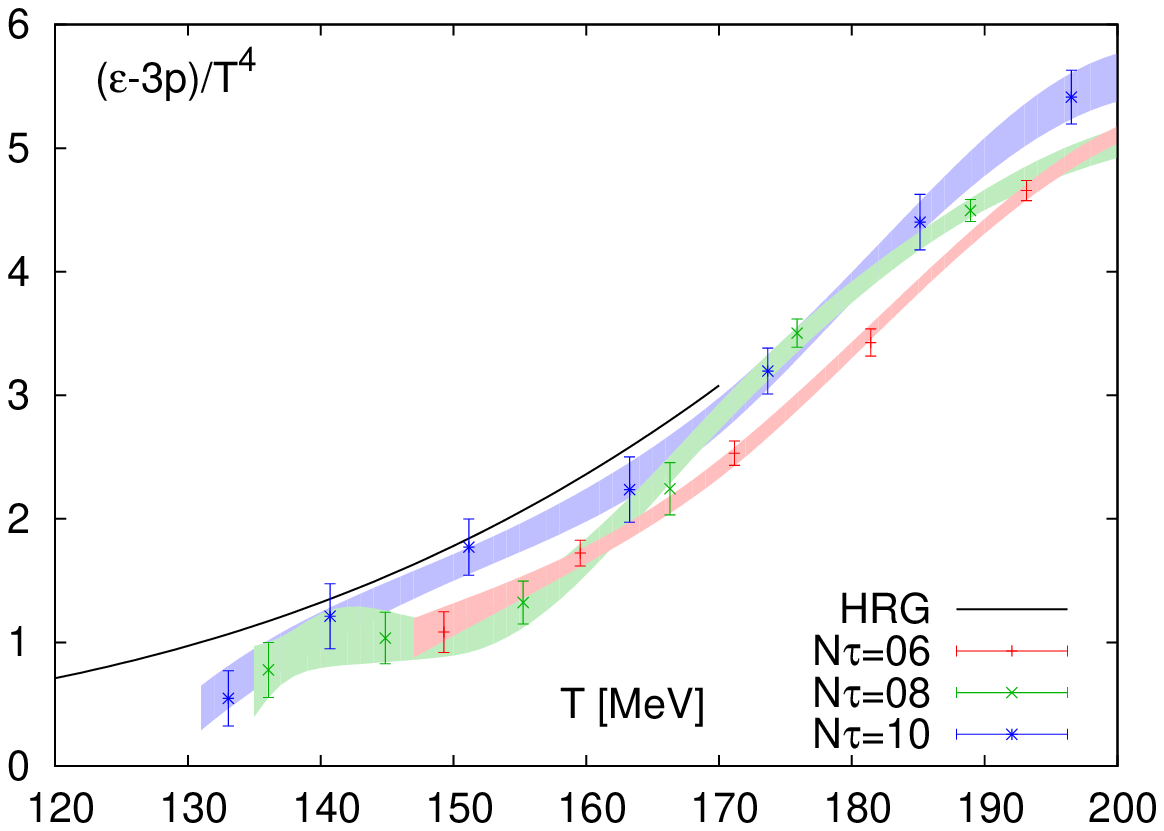}\hfill
\includegraphics[width=0.495\textwidth]{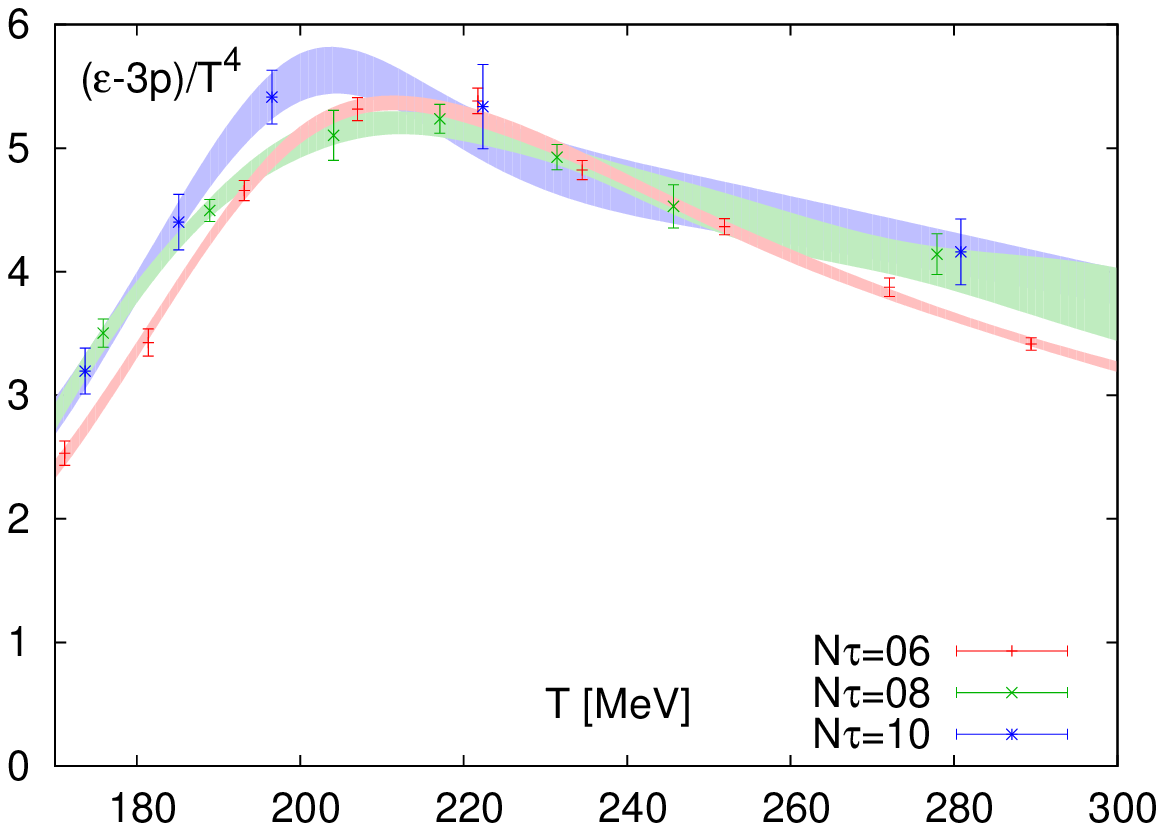}
\end{center}
\vspace{-7mm}
\caption{The trace anomaly at low temperatures (left) and in the peak
region (right). Bands represent statistical errors on the spline fits.}
\vspace{-3mm}
\label{fig_low_peak}
\end{figure}

The contribution due to the variation of the charm quark mass in the Naik term,
$dM_c/d\epsilon_N$ is larger on coarser lattices. We have evaluated it on
$N_\tau=6$ lattices at $T=272$ and $369$~MeV, and corresponding zero-temperature
ensembles. This quantity is about 2\% of the total result, which is comparable to
the statistical errors. Therefore in the present analysis, in particular, for the
quantity in Fig.~\ref{fig_Sgf} (bottom right) and for the total trace anomaly, we did
not include the term $\Delta(\bar\psi_c [dM_c/d\varepsilon_N]\psi_c)$.
Its inclusion will presumably decrease the cutoff effects on the charm contribution
at low temperatures, $150-200$~MeV.

We fitted the data for the trace anomaly with splines, shown as bands
in Fig.~\ref{fig_low_peak}. The width of the bands represents only the statistical error,
estimated by bootstrap. In Fig.~\ref{fig_low_peak} (left) the low-temperature
region is shown. The solid line is the hadron resonance gas (HRG) model result.
Coarser lattices produce a heavier hadron spectrum, thus, the approach
to the continuum is from below (at least, when the temperature is set 
with the $r_1$ scale). In the peak region, Fig.~\ref{fig_low_peak} (right), the cutoff
effects are mild, and at temperatures around 200~MeV and above, where the light
hadrons melt, they are presumably not caused by taste symmetry breaking in the quark
sector.

\begin{figure}
\begin{center}
\includegraphics[width=0.85\textwidth]{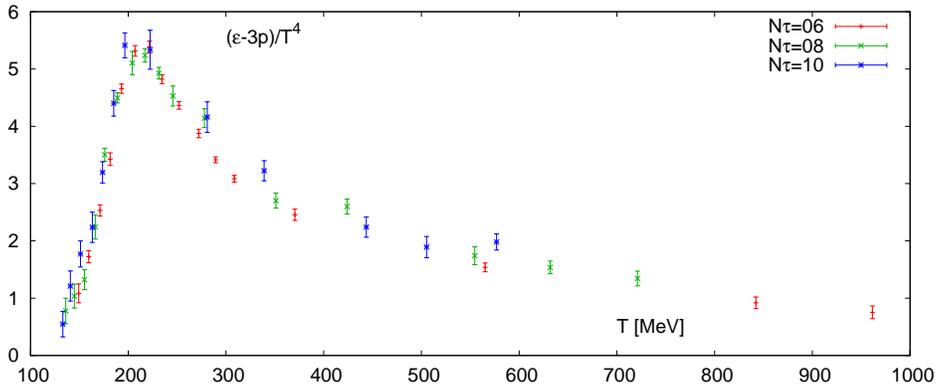}
\end{center}
\vspace{-7mm}
\caption{The 2+1+1 flavor trace anomaly with HISQ along the $m_l=m_s/5$ LCP.}
\vspace{-3mm}
\label{e3p_wide}
\end{figure}

The total trace anomaly is shown in Fig.~\ref{e3p_wide}. The $N_\tau=10$ lattices
extend to 580~MeV. This is enough to cover the peak of the valence charm 
contribution, but without further $N_\tau=10$ data the cutoff effects above
this temperature are thus hard to quantify.

\section{Conclusion}

We have extended our calculation of the 2+1+1 flavor QCD equation of state
with highly improved staggered quarks in two ways. We generated several new
zero- and finite-temperature ensembles to provide better coverage of temperatures
in the range 130-1000~MeV and substantially increased the statistics on most
of the finite temperature ensembles. We have reached lattice spacings down
to $0.034$~fm, which corresponds to 967~MeV on the coarsest, $N_\tau=6$ lattice.
The charm contribution to the trace anomaly becomes non-negligible around 300~MeV
and reaches the maximum far in the deconfined phase, around 500~MeV.
The cutoff effects on the trace anomaly are significant at low temperatures
and are mild in the peak region. However, $N_\tau=12$ ensembles will be needed
for a reliable continuum extrapolation.

\section*{Acknowledgments}

This work was supported by the U.S. Department of Energy and National Science Foundation.
Computations for this work were done at the Argonne Leadership Computing Facility (ALCF), the National Center for Atmospheric Research (UCAR), Bluewaters at the National Center for Supercomputing Resources (NCSA), the National Energy Resources Supercomputing Center (NERSC), the National Institute for Computational Sciences (NICS), the Texas Advanced Computing Center (TACC), and the USQCD facilities at Fermilab, under grants from the NSF and DOE.

\end{document}